\documentclass[seceq]{ptptex}
\usepackage{amsmath,amssymb}
\usepackage{graphicx}
\usepackage{booktabs}

\def\boldv{\mbox {\boldmath{$v$}}}
\def\m2{\hspace*{-0.2cm}}

\def\th{\theta}

\def\bD{\breve{D}}

\def\cL{{\cal L}}
\def\cM{{\cal M}}

\def\half{{1 \over 2}}

\def\sqr2{\sqrt{2}}

\title{%        %You can use \\ for explicit line-break
Dirac Mass Matrices in Gauge Field Theory\\ of Horizontal Symmetry
}

\author{%       %Use \sc for the family name
Yasufumi {\sc Konishi}\footnote{E-mail address: konishi@cc.kyoto-su.ac.jp}
and Ikuo S. {\sc Sogami}\footnote{E-mail address: sogami@cc.kyoto-su.ac.jp}}

\inst{%         %Affiliation, neglected when [addenda] or [errata]
Physics Department, Kyoto Sangyo University, Kyoto 603-8555, Japan
}

\recdate{%      %Editorial Office will fill in this.
\today
}

\begin{document}

\abst{%         %this abstract is neglected when [addenda] or [errata]
We investigate Dirac mass matrices derived in the gauge field theory
of a horizontal symmetry generated by a central extension of the
Pauli algebra. Through numerical analyses of the observed data of the
charged fermion masses and the flavor mixing matrix of quarks, values of
free parameters in the mass matrices are determined and several empirical
relations are found among the Yukawa coupling constants. As one specific
feature of the theory, we find different orderings in squared mass eigenvalues
for the up and down quark sectors.
}

\maketitle

\section{Introduction}
The Standard Model (SM) of particle physics possesses no principle
to lay any restriction on the pattern of the Yukawa interactions.
Nine complex coupling constants are treated as free parameters in
every four sectors consisting of three generations of quarks and
leptons. One way to describe order and variety of
the generational structure is to postulate a gauge symmetry called
the {\it horizontal symmetry}.~\cite{Wilczek.1979, Shaw.1993}
In the previous paper,~\cite{Sogami.2009} one of the present authors
has proposed a gauge field theory of a new horizontal symmetry.
The purpose of this paper is to investigate the Dirac mass matrices
derived in the theory and make numerical analyses of the mass spectra of
the charged fermions and the flavor mixing matrix (FMM) of quarks.

The horizontal (H) symmetry of the theory~\cite{Sogami.2009} is
postulated to be described by the Lie group generated by a central
extension of the Pauli algebra which was found in investigating the FMM
of quarks and leptons~\cite{Sogami,SogamiKon}. The algebra
consisting of four generators has the central element identified with
the democratic matrix\cite{DemocraticMM} which can create hierarchical
mass spectra of fundamental fermions. The number of the Yukawa coupling
constants of the theory is reduced to $4/9$ of that of the SM. Through
spontaneous breakdown of the symmetry, the theory leads to the Dirac
mass matrices $\cM_f$ for the sector $f$ with definite electric charge.

The matrix $\cM_f$ with unique non-Hermitian structure possesses four
unknown complex parameters. To deduce information on the $f$ sector, we must
solve the eigenvalue problem of the Hermitian matrix $\cM_f\cM_f^\dagger$.
It should be noticed that unspecified quantities involved in these Hermitian
matrices are reduced down to ten real parameters. Consequently, numerical
analysis on the experimental data~\cite{Amsler.2008} on the six masses
and four FMM parameters of the quark sector enables us to determine the
values of ten real parameters, from which we find several empirical relations
among the Yukawa coupling constants.

The Dirac mass matrices deduced from the Lagrangian density of the Yukawa
interactions are shown explicitly in \S 2. We describe formalisms
for the eigenvalue problem of $\cM_f\cM_f^\dag$ in \S 3 and \S 4. 
Different mass orderings in the up and down quark sectors are explained
in \S 5. In \S 6, values of the parameters in the mass matrices are
determined by numerical analysis and specific empirical relations are
found among the Yukawa coupling constants. Discussion is given in \S 7
and functional dependence among parameters and mass eigenvalues is
examined in Appendix.

\section{Dirac mass matrices}
In the low energy region less than the electroweak scale
$v=246$\,{GeV},\cite{Djouadi.2008}
the gauge field theory of the H symmetry provides the effective
theory for flavor phenomenology. Breakdowns of the H and electroweak
symmetries lead to the Lagrangian density for Dirac masses of
the quarks and leptons in the following forms~\cite{Sogami.2009}
\begin{equation}
 \cL^{\rm Y}_{\cM}
 =  \sum_{f=u,d}\,\bar{\Psi}^{\,f}_L\cM_f\Psi^{\,f}_R
  + \sum_{f=\nu,e}\,\bar{\Psi}^{\,f}_L\cM_f\Psi^{\,f}_R
  + {\rm h.c.}
\end{equation}
in which $\Psi^{f}_{L,R}$ are chiral fermion fields and $\cM_f$ is the Dirac
mass matrix. For the up sectors ($f=u,\,\nu$) of the electroweak symmetry,
the mass matrix is given by
\begin{equation}
   \cM_{f} = a_f\,I
             + \frac{1}{\sqrt{3}}
               b_{f1}\left(
                      \begin{array}{rrr}
                         1 &  1 &  1  \\
                        -1 & -1 & -1 \\
                         0 &  0 &  0 \\
                      \end{array}
                     \right)
             + \frac{1}{3}
                  b_{f2}\left(
                       \begin{array}{ccc}
                         -1 & -1 & 2 \\
                         -1 & -1 & 2 \\
                         -1 & -1 & 2 \\
                       \end{array}
                     \right)
                     + c_f\,\bD
\label{EWUpDirac}
\end{equation}
where
\begin{equation}
    \bD = \frac{1}{3}
        \left(
         \begin{array}{ccc}
           1 & 1 & 1\\
           1 & 1 & 1\\
           1 & 1 & 1\\
         \end{array}
        \right)
\end{equation}
is the democratic element.\cite{DemocraticMM} The four coefficients in the mass
matrix are expressed in terms of the Yukawa coupling constants $Y_{fi}$ and
the vacuum expectation value $v$ of the scalar field as $a_f = Y_{f1}\,v$,
$b_{f1}= -Y_{f2}\,v$, $b_{f2} =Y_{f3}\,v$ and $c_f= 3Y_{f4}\,v$. For the down
sectors ($f=d,\,e$), we find
\begin{equation}
   \cM_{f} = a_f\,I
             + b_{f1}\left(
                      \begin{array}{ccc}
                        0 & 0 & 0 \\
                        0 & 0 & 0 \\
                        1 & 1 & 1 \\
                      \end{array}
                     \right)
             + \frac{1}{\sqrt{3}}
                b_{f2}\left(
                       \begin{array}{ccc}
                         1 & -1 & 0 \\
                         1 & -1 & 0 \\
                         1 & -1 & 0 \\
                       \end{array}
                     \right)
              + c_f\,\bD
\label{EWDownDirac}
\end{equation}
where $a_f = Y_{f1}\,v$, $\ b_{f1}= Y_{f2}\,v$, $\ b_{f2}= Y_{f3}\,v$
and $c_f = 3Y_{f4}\,v$.

In order to diagonalize the non-Hermitian mass matrix $\cM_f$, it is
necessary to have recourse to the bi-unitary transformation~\cite{ChengLi}
\begin{equation}
        V_L^{f\dag}\cM_f V_R^{f} = \cM_{f{\rm diagonal}}.
\end{equation}
To derive the mass eigenvalues and the diagonalizing matrix $V_{L}^{f}$,
it is necessary to solve the eigenvalue problem for the self-adjoint matrix
$\cM_f\cM_f^{\dagger}$ as
\begin{equation}
 \cM_f\cM^{\dag}_f\,|\boldv^{(f)i}\rangle = m_i^{(f)2}\,|\boldv^{(f)i}\rangle
 \label{mass2eigenvalue}
\end{equation}
for each charged fermion sector ($f=u, d, e$). The diagonalizing matrix
$V_{L}^{f}$ is obtained in terms of the eigenvectors $|\boldv^{(f)i}\rangle$
and the FMM of the quark sector is constructed by
\begin{equation}
   V = V_L^{u\dag}V_L^{d}
     = \left(\,\langle \boldv^{(u)i}|\boldv^{(d)j}\rangle\,\right),
\end{equation}
provided that the eigenvectors are arranged in increasing orders of the
masses for up and down sectors.

\section{Eigenvalue problem 1: Quark FMM}
To solve the eigenvalue problem for $\cM_f\cM^{\dag}_f$, it turns out
convenient to use the basis vectors 
\begin{equation}
  \begin{array}{lll}
      |\,1\ \rangle
        = \frac{1}{\sqrt{2}}
         \left(
           \begin{array}{r}
               1\\
           \noalign{\vskip 0.05cm}
              -1\\
           \noalign{\vskip 0.05cm}
               0
           \end{array}
         \right),
     &
       |\,2\ \rangle
         = \frac{1}{\sqrt{6}}
          \left(
            \begin{array}{r}
                1\\
            \noalign{\vskip 0.05cm}
                1\\
            \noalign{\vskip 0.05cm}
               -2
            \end{array}
          \right),
     &
       |\,3\ \rangle
         = \frac{1}{\sqrt{3}}
          \left(
            \begin{array}{r}
                1\\
            \noalign{\vskip 0.05cm}
                1\\
            \noalign{\vskip 0.05cm}
                1
            \end{array}
          \right)
  \end{array} ,
  \label{eigenvectors}
\end{equation}
which are eigenvectors of the democratic element $\bD$. With these bases,
the eigenvector in (\ref{mass2eigenvalue}) is expanded as
\begin{equation}
   |\,\boldv^{(f)}\,\rangle
    = x_{f}|\,1\,\rangle + y_{f}|\,2\,\rangle + z_{f}|\,3\,\rangle.
\end{equation}

For the up quark sector, (\ref{mass2eigenvalue}) is rewritten for the
coefficients of $|\boldv^{(u)}\rangle$ as
\begin{equation}
 \left(
  \begin{array}{ccc}
   |a_u|^2 + 2|b_{u1}|^2   &       0              &  \sqrt{2}C_u^\ast b_{u1}\\
   \noalign{\vskip 0.2cm}
     0                     &       |a_u|^2        & -\sqrt{2}a_u b_{u2}^\ast\\
   \noalign{\vskip 0.2cm}
   \sqrt{2}C_u b_{u1}^\ast &-\sqrt{2}a_u^\ast b_{u2} & |C_u|^2 + 2|b_{u2}|^2
  \end{array}
 \right)
 \left(
  \begin{array}{c}
     x_u\\
   \noalign{\vskip 0.2cm}
     y_u\\
   \noalign{\vskip 0.2cm}
     z_u\\
  \end{array}
 \right)
 =
  m^{(u)2}\,
  \left(
  \begin{array}{c}
     x_u\\
    \noalign{\vskip 0.2cm}
     y_u\\
    \noalign{\vskip 0.2cm}
     z_u\\
  \end{array}
 \right)
\label{MMdup123base}
\end{equation}
where $C_u= c_u + a_u$. Similarly, for the down quark sector, we obtain
\begin{equation}
 \left(
  \begin{array}{ccc}
   |a_d|^2                &         0               & \sqrt{2}a_db_{d2}^\ast \\
   \noalign{\vskip 0.2cm}
     0                    & |a_d|^2 + 2|b_{d1}|^2   &-\sqrt{2}C_d^\ast b_{d1}\\
   \noalign{\vskip 0.2cm}
   \sqrt{2}a_d^\ast b_{d2} & -\sqrt{2}C_db_{d1}^\ast &  |C_d|^2 + 2|b_{d2}|^2
  \end{array}
 \right)
 \left(
  \begin{array}{c}
     x_d\\
   \noalign{\vskip 0.2cm}
     y_d\\
   \noalign{\vskip 0.2cm}
     z_d\\
  \end{array}
 \right)
 =
  m^{(d)2}
  \left(
  \begin{array}{c}
     x_d\\
    \noalign{\vskip 0.2cm}
     y_d\\
    \noalign{\vskip 0.2cm}
     z_d\\
  \end{array}
 \right)
\label{MMddown123base}
\end{equation}
for the coefficients of $|\boldv^{(d)}\rangle$, where $C_d= c_d+a_d+b_{d1}$.

To clarify the counting of independent parameters in these equations and the
FMM, we define phase factors by
\begin{equation}
 C_u^\ast b_{u1} = |C_u b_{u1}|e^{i \mu_u},
~
 a_u b_{u2}^\ast = |a_u b_{u2}|e^{i \nu_u},
~
 a_d b_{d2}^\ast = |a_d b_{d2}|e^{i \mu_d},
~
 C_d^\ast b_{d1} = |C_d b_{d1}|e^{i \nu_d}
\end{equation}
and introduce the diagonal phase matrix
\begin{equation}
 P^f = {\rm diag}(\, e^{i \mu_f},\   e^{i \nu_f},\  1\,)
\end{equation}
for the $f$ sector to adjust phase factors. By separating the diagonal phase
matrices, the eigenvectors of (\ref{MMdup123base}) and (\ref{MMddown123base})
are found, respectively, in the forms
\begin{equation}
   P^u {\boldsymbol{u}}_j^u
     = P^u N_j^u
     \left(
       \begin{array}{c}
         \sqrt{2}|C_u b_{u1}|(m_j^{(u)2} - |a_u|^2 )\\
         \noalign{\vskip 0.2cm} 
          - \sqrt{2}|a_ub_{u2}|(m_j^{(u)2} - |a_u|^2 - 2|b_{u1}|^2 )\\
         \noalign{\vskip 0.2cm} 
         (m_j^{(u)2} - |a_u|^2 )(m_j^{(u)2} - |a_u|^2 - 2|b_{u1}|^2 )
       \end{array}
     \right)
     \label{exactupeigenvectors}
\end{equation}
and
\begin{equation}
   P^d {\boldsymbol{u}}_j^d 
      = P^d N_j^d
      \left(
        \begin{array}{c}
            \sqrt{2}|a_db_{d2}|(m_j^{(d)2} - |a_d|^2 - 2|b_{d1}|^2 )\\
        \noalign{\vskip 0.3cm}
          - \sqrt{2}|C_d b_{d1}|(m_j^{(d)2} - |a_d|^2 )\\
        \noalign{\vskip 0.3cm}
           (m_j^{(d)2} - |a_d|^2 )(m_j^{(d)2} - |a_d|^2 - 2|b_{d1}|^2 )
        \end{array}
      \right)
      \label{exactdowneigenvectors}
\end{equation}
where $m_j^{(f)2}$ are eigenvalues of squared masses and $N_j^f$ are
the normalization constants.
Then, with the orthogonal matrices
\begin{equation}
   O_L^f = \left({\boldsymbol{u}}^{f}_1,\ {\boldsymbol{u}}^{f}_2,
   \ {\boldsymbol{u}}^{f}_3 \right)
   \label{orthogonal}
\end{equation}
consisting of the vectors ${\boldsymbol{u}}_j^f$, the FMM for the quark sector
is calculated to be
\begin{equation}
   V = O_L^{u\dag}PO_L^{d}
\end{equation}
with the diagonal phase matrix
\begin{equation}
  P = {\rm diag}(\, e^{i \mu},\  e^{i \nu},\  1\,) = P^{u\dag}P^{d},
  \quad \mu=\mu_d - \mu_u, \quad \nu=\nu_d - \nu_u .
\end{equation}

This FMM includes unknown parameters of eight real numbers and two phases.
The secular equations for the eigenvalue problems in (\ref{MMdup123base})
and (\ref{MMddown123base}) work to fix six real parameters in terms of the
mass eigenvalues. Consequently, two real numbers and two phases remain
unspecified in the FMM for the quark sector.

\section{Eigenvalue problem 2: Mass spectra of charged fermions}
Since the secular equations for both of the eigenvalue problems in
(\ref{MMdup123base}) and (\ref{MMddown123base}) take the same form,
the suffix $f$ is omitted for all quantities in this section and the
Appendix. In terms of the shifted variable $s = m^2 - |a|^2$, the secular
equation is obtained as
\begin{equation}
  s^3 -(|C|^2 - |a|^2 + 2|b|^2)\,s^2
  -2(|a b|^2 - 2|b_{1}b_{2}|^2 )\,s + 4|a b_{1} b_{2}|^2 = 0
\end{equation}
where
\begin{equation}
 |b|^2=|b_1|^2+|b_2|^2 .
\end{equation}
Let us solve this equation by
the Cardano method. Introducing the dimensionless quantities
\begin{equation}
   P = \frac{1}{3}
       \frac{|C|^2 - |a|^2 + 2|b|^2}{|ab_1b_2|^{\frac{2}{3}}},\ \ 
   Q = \frac{2}{3}
       \frac{|a|^2|b|^2 - 2|b_1 b_2|^2}{|ab_1b_2|^{\frac{4}{3}}},
\end{equation}
and changing the variable by $s = \left|ab_1b_2\right|^{\frac{2}{3}}(t+P)$,
we obtain the reduced proper equation without the second order term as follows:
\begin{equation}
   t^3 -3(P^2 + Q)t -2P^3 -3PQ +4= 0 .
   \label{ReducedEq}
\end{equation}
One solution of this equation is determined as $t=t_+ +t_-$ by the sum of
two quantities $t_+$ and $t_-$ which are subject to the relations
\begin{equation}
  t_{\pm}^3 = \half\left(2P^3 + 3PQ -4 \pm i\sqrt{|D|}\right)
\end{equation}
where
\begin{equation}
  D = -16P^3 - 3P^2Q^2 - 24PQ - 4Q^3 + 16 .
\end{equation}
In the analysis below, it is appropriate to use the polar representation
$t_+ = \rho{\rm e}^{i|\th|}$ and $t_- = \rho{\rm e}^{-i|\th|}$ in which
$\rho$ and $\theta$ are expressed, in terms of $P$ and $Q$, as
\begin{equation}
  \rho = \sqrt{P^2 + Q},\quad
  \tan 3|\th| = \frac{\sqrt{|D|}}{2P^3 + 3PQ -4} .
\label{rhotan3theta}
\end{equation}
Then, the three solutions of (\ref{ReducedEq}) are derived to be
\begin{equation}
 \begin{array}{lll}
  t_1 \!\!\!&=\omega t_+ +\omega^2 t_- \!\!&= 2\rho\cos(|\th|+\frac{2\pi}{3})\\
  \noalign{\vskip 0.3cm}
  t_2 \!\!\!&=\omega^2 t_+ +\omega t_- \!\!&= 2\rho\cos(|\th|+\frac{4\pi}{3})\\
  \noalign{\vskip 0.3cm}
  t_3 \!\!\!&= t_+ + t_-               \!\!&= 2\rho\cos\,\th
 \end{array}
\end{equation}
where $\omega=\exp(i2\pi/3)$.

In this way, we have solved the eigenvalue problems in (\ref{MMdup123base}) and
(\ref{MMddown123base}) obtaining the squared masses as follows:
\begin{equation}
 \begin{array}{l}
  m_{1}^2 = |a|^2 + \frac{1}{3}\left( |C|^2 - |a|^2 + 2|b|^2 \right)
             \left[ 1 + 2\sqrt{1+\delta}\cos(|\th|+\frac{2\pi}{3})\right] ,\\
  \noalign{\vskip 0.3cm}
  m_{2}^2 = |a|^2 + \frac{1}{3}\left( |C|^2 - |a|^2 + 2|b|^2 \right)
             \left[ 1 + 2\sqrt{1+\delta}\cos(|\th|+\frac{4\pi}{3})\right] ,\\
  \noalign{\vskip 0.3cm}
  m_{3}^2 = |a|^2 + \frac{1}{3}\left( |C|^2 - |a|^2 + 2|b|^2 \right)
             \left[ 1 + 2\sqrt{1+\delta}\cos\th \right] ,
 \end{array}
 \label{exactmassfomulae}
\end{equation}
where the parameter
\begin{equation}
  \delta= 6\frac{|a|^2|b|^2 - 2|b_{1}b_{2}|^2}
                   {[|C|^2 - |a|^2 + 2|b|^2]^2}
\label{delta.def}
\end{equation}
is introduced to simplify the expressions.

As confirmed below, the magnitude of the angle $|\th|$ must be
sufficiently small for the mass spectra to have hierarchical structure.
Here, it is crucially important to note that the squared masses have
orderings $m_1^2 < m_2^2 < m_3^2$ and $m_2^2 < m_1^2 < m_3^2$, respectively,
for $\th>0$ and $\th<0$.

The squared masses in (\ref{exactmassfomulae}) depend on the four real
quantities $|a|^2$, $|b_1|^2$, $|b_2|^2$ and $|C|^2$. In the present analysis,
we interpret inversely that $|b_1|^2$, $|b_2|^2$ and $|C|^2$ are functions
of the masses and the parameter $|a|^2$. Then, all quantities for the quark
FMM are determined in terms of the observed quark masses and the independent
parameter $|a|$. (See the Appendix).

In the following sections, we make numerical analysis of the quark FMM
by using experimental values of quark masses as inputs and by adjusting
the independent parameters $|a_u|$ and $|a_d|$.

\section{Mass orderings of the up and down quark sectors}\label{Massordering}
For numerical analyses below, we use the quark masses at the energy
scale of the $Z$ boson, i.e., $m_Z=91.2$\,GeV. The values calculated
by the renormalization group equations\cite{Fusaoka.1998} are given
as follows:\cite{Xing.2008}
\begin{equation}
\begin{array}{lll}
 m_u = 1.27^{+0.50}_{-0.42} \, {\rm MeV} ,&~~
 m_c = 0.619 {\pm 0.084} \, {\rm GeV} ,&~~
 m_t = 171.7 {\pm 3.0} \, {\rm GeV} ,\\
 \noalign{\vskip 0.2cm}
 m_d = 2.90^{+1.24}_{-1.19} \, {\rm MeV} ,&~~
 m_s = 55^{+16}_{-15} \, {\rm MeV} ,&~~
 m_b = 2.89 {\pm 0.09} \, {\rm GeV} .
\end{array} 
\label{quarkmassdata}
\end{equation}

The observed FMM of the quark sector has the prominent feature that the
matrix elements decrease rapidly for each step away from the diagonal.
To reproduce such characteristics, both of the orthogonal matrices $O_L^u$
and $O_L^d$ in (\ref{orthogonal}) must approximately be close to the unit
matrix. Accordingly, as a step for the FMM analysis, it is reasonable to
examine numerically which of the solutions characterized by $\theta>0$ and
$\theta<0$ in (\ref{exactmassfomulae}) and the associated eigenvectors in
(\ref{exactupeigenvectors}) and (\ref{exactdowneigenvectors}) can bring
the orthogonal matrix nearer to the unit matrix.

Let us apply the positive-$\theta$ solution with the experimental mass values
in (\ref{quarkmassdata}) to examine the orthogonal matrices. For the down
quark sector, it is proved that all of the diagonal elements of the matrix
$O_L^d$ can approach to the unit for small value of $|a_d|$. Contrastingly,
for the up quark sector, some of the diagonal elements of the matrix $O_L^u$
is shown to be quite smaller than 1 for any value of $|a_u|$. The situation
reverses completely for the negative-$\theta$ solution. Numerical calculations 
with the negative-$\theta$ solution show that, while $O_L^u$ approaches to
the unit matrix by adjusting $|a_u|$, $O_L^d$ can not be made close to the unit
matrix for any value of $|a_d|$.

Accordingly, it is necessary to choose the positive- and negative-$\theta$
solutions, respectively, for the down and up quark sectors to reproduce
the experimental results of the quark FMM. The masses and state vectors
of the observed down quark members ($d,\,s,\,b$) must be described by the
positive-$\theta$ solution with normal ordering as follows:
\begin{equation}
 \begin{array}{ccc}
  m_d=m_1^{(d)}, & m_s=m_2^{(d)}, & m_b=m_3^{(d)} ; \\
  \noalign{\vskip 0.2cm}
  |\boldv_d{}\rangle=|\boldv^{(d)1}{}\rangle, & 
  |\boldv_s{}\rangle=|\boldv^{(d)2}{}\rangle, & 
  |\boldv_b{}\rangle=|\boldv^{(d)3}{}\rangle .
 \end{array}
 \label{normal}
\end{equation}
As for the observed up quark members ($u,\,c,\,t$), their masses and state
vectors have to be identified with the quantities of the negative-$\theta$
solutions with partly-reversed ordering as follows:
\begin{equation}
 \begin{array}{ccc}
  m_u=m_2^{(u)}, & m_c=m_1^{(u)}, & m_t=m_3^{(u)} ; \\
  \noalign{\vskip 0.2cm}
  |\boldv_u{}\rangle=|\boldv^{(u)2}{}\rangle, & 
  |\boldv_c{}\rangle=|\boldv^{(u)1}{}\rangle, & 
  |\boldv_t{}\rangle=|\boldv^{(u)3}{}\rangle .
 \end{array}
 \label{reversed}
\end{equation}
In the present theory, it is crucial to accept these interpretations of
the solutions of the eigenvalue problems in (\ref{mass2eigenvalue}).

\section{Hierarchical structure of the Yukawa coupling constants}
As shown in the Appendix, the quantities $|b_{f1}|$, $|b_{f2}|$ and $|C_{f}|$
are expressed in terms of the masses and the adjustable parameter $|a_{f}|^2$
in the hierarchical approximation ($m_3^{(f)2} \gg m_1^{(f)2}, m_2^{(f)2}$).
Using these results and accepting the interpretations in (\ref{normal}) and
(\ref{reversed}), we make numerical analysis of the quark FMM. The best fit
to the FMM data of the Particle Data Group\cite{Amsler.2008} is obtained
with the following values of the four parameters as
\begin{equation}
 |a_u|= 30.4 , \quad
 |a_d|= 13.2 , \quad
 \mu= 0.96 , \quad
 \nu=2.32 
\end{equation}
which lead to the magnitude of the elements of the quark FMM as
\begin{equation}
 |V| =
 \left(
  \begin{array}{ccc}
       0.974210  	&       0.225705 	&       0.003595 \\
   \noalign{\vskip 0.1cm}
       0.225473 	&       0.973343 	&       0.041511  \\	
   \noalign{\vskip 0.1cm}
       0.008723 	&       0.040746 	&       0.999132  \\
  \end{array}
 \right),
\end{equation}
and the Jarlskog invariant measure\cite{Jarlskog.1985}
\begin{equation}
   J = 3.1 \times 10^{-5} 
\end{equation}
for the CP violation. The values of all parameters for the best fit found
above are listed in Table \ref{10parameters}.
\begin{table}[htbp]
\caption{Values of ten parameters}
\begin{center}
\begin{tabular}{lll}
\toprule
up quark (MeV) & down quark (MeV) & \ phases\\
\midrule
$|a_u|$ \ =  30.4 &
$|a_d|$ \ =  13.2 &
$\mu$   =   0.96 \\
\noalign{\vskip 0.1cm}
$|b_{u1}|$ =  831  &
$|b_{d1}|$ =  92.1 &
$\nu$      =   2.32\\
\noalign{\vskip 0.1cm}
$|b_{u2}|$ =  63800 &
$|b_{d2}|$ =    818 &\\
\noalign{\vskip 0.1cm}
$|C_u|$ =  146000 &
$|C_d|$ =    2650 &\\
\bottomrule
\end{tabular}
\label{10parameters} 
\end{center}
\end{table}

Results in Table \ref{10parameters} show clearly that the real parameters
satisfy the hierarchical orderings
$|a_f|^2 \ll |b_{f1}|^2 \ll |b_{f2}|^2 \ll |C_f|^2$
for each quark sector. By making more careful comparison among them, we find 
approximate relations
\begin{equation}
\frac{|b_{u1}|}{|b_{d2}|} \sim 1, ~~
\frac{|b_{d2}|}{|b_{d1}|} \sim 9^1, ~~
\frac{|b_{u2}|}{|b_{u1}|} \sim 9^2
\label{PhenoRelationConstants}
\end{equation}
which result readily in interesting empirical relations
\begin{equation}
\frac{|Y_{u2}|}{|Y_{d3}|} \sim 1, ~~
\frac{|Y_{d3}|}{|Y_{d2}|} \sim 9^1, ~~
\frac{|Y_{u3}|}{|Y_{u2}|} \sim 9^2, 
\label{PhenoRelationYukawa}
\end{equation}
among the four Yukawa coupling constants.

The largest quantity in Table \ref{10parameters} is $|C_f|$ for both
quark sectors. Note that the original parameter $c_f$ in the mass matrix
$\cM_f$ can appear only through the quantity $C_f$, as $C_u=c_u+a_u$ and
$C_d=c_d+a_d+b_{d1}$, in the Hermitian matrix $\cM_f\cM_f^\dag$. Due to
this feature which works to decrease the number of unknown quantities
in $\cM_f\cM_f^\dag$, we can do nothing but determine $|c_f|$ approximately
$|c_u| \approx |C_u|$ and $|c_d| \approx |C_d|$.

Using $v=246$\,GeV and the data in Table \ref{10parameters}, we are able to
fix the magnitudes of the Yukawa coupling constants of the quark sectors as
in Table \ref{YCCvalues}. In the somehow crude approximation for $|c_f|$,
the constant $|Y_{f4}|$ have comparatively large uncertainty.

\begin{table}[hbp]
 \begin{center}
\caption{Yukawa coupling constants}
  \begin{tabular}{cccc}
\toprule
 $f$ & $u$ & $d$ & $e$ \\
\midrule
$|Y_{f1}|$ & $1.2\times 10^{-4}$ & $5.4\times 10^{-5}$ & $3.5\times 10^{-5}$ \\
\noalign{\vskip 0.1cm}
$|Y_{f2}|$ & $3.4\times 10^{-3}$ & $3.7\times 10^{-4}$ & $4.1\times 10^{-4}$ \\
\noalign{\vskip 0.1cm}
$|Y_{f3}|$ & $2.6\times 10^{-1}$ & $3.3\times 10^{-3}$ & $3.6\times 10^{-3}$ \\
\midrule
$|Y_{f4}|$ & $2.0\times 10^{-1}$ & $3.6\times 10^{-3}$ & $1.6\times 10^{-3}$ \\
\bottomrule
\label{YCCvalues}
  \end{tabular}
 \end{center}
\end{table}

Thus far, the Yukawa coupling constants of quark sectors are calculated
so as to recreate the observed data of the quark FMM. This approach is
not applicable, as it stands, to the lepton sector. Here, let us find an
approximate scheme with less numbers of adjustable parameters based on the
benefit of hindsight on the empirical relations in (\ref{PhenoRelationYukawa})
and apply it to analyze the masses of charged leptons.

For its purpose, we introduce a new variable $\beta$ by the equations
\begin{equation}
 |b_{d1}|= 9\beta,\quad |b_{d2}|= |b_{u1}|= 9^2\beta,\quad |b_{u2}|= 9^4\beta,
\label{approximateassmpquark}
\end{equation}
which verify all of the relations in (\ref{PhenoRelationConstants}), and
reinvestigate the quark masses and FMM in terms of seven quantities
$|a_u|, |a_d|, |C_u|, |C_d|, \mu, \nu$ and $\beta$. 
\begin{table}[htbp]
 \begin{center}
 \caption{Values of parameters for approximate estimation}
  \begin{tabular}{lll}
\toprule
 up quark (MeV) & down quark (MeV) & common parameters\\
\midrule
$|a_u|$ =          30 & $|a_d|$ =  13   & $\mu$ = 0.95,\ \  $\nu$ = 2.3\\
\noalign{\vskip 0.1cm}
$|C_u|$ =      142000 & $|C_d|$ =  2620 & $\beta$ = 10\\
\bottomrule
\label{7parameters}
  \end{tabular}
 \end{center}
\end{table}
It turns out possible
to explain all observed data within the range of experimental errors.
In fact, by using the values of seven quantities in Table \ref{7parameters},
we obtain
\begin{equation}
 \begin{array}{lll}
 m_u = 1.20 \, {\rm MeV} ,&~~
 m_c = 0.628 \, {\rm GeV} ,&~~
 m_t = 169.6 \, {\rm GeV} , \\
 m_d = 2.87 \, {\rm MeV} ,&~~
 m_s = 53.8 \, {\rm MeV} ,&~~
 m_b = 2.86 \, {\rm GeV} ,
 \end{array}
\end{equation}
for the six quark masses and
\begin{equation}
 |V| =
 \left(
  \begin{array}{ccc}
       0.973984 	&       0.226589 	&       0.003509  \\
   \noalign{\vskip 0.1cm}
        0.226452 	&          0.973161 	&       0.040962  \\
   \noalign{\vskip 0.1cm}
       0.008616 	&       0.040199 	&       0.999155  \\
  \end{array}
 \right)
,~~~
J=       2.9 \times 10^{-5} 
\end{equation}
for the magnitudes of the quark FMM elements and the Jarlskog parameter.

For both sectors of the down quark and the charged lepton, the Dirac mass
matrices take the common forms in (\ref{EWDownDirac}). To analyze the masses
of the charged leptons in this approximate scheme, we assume that the
empirical equations related to the down quarks in (\ref{PhenoRelationConstants})
hold also in the charged lepton sector. Then, following the equations
in (\ref{approximateassmpquark}), we introduce a new variable $\beta_e$ by
the relations
\begin{equation}
 |b_{e1}| = 9\beta_e, \quad |b_{e2}| = 9^2\beta_e
\label{approximateassmplepton}
\end{equation}
and analyze the charged lepton masses in terms of the three parameters $|a_e|$,
$|C_e|$ and $\beta_e$. Numerical estimation shows readily that the charged
lepton masses\cite{Fusaoka.1998,Xing.2008}
\begin{equation}
 m_e = 0.4866 \, {\rm MeV}, \quad
 m_{\mu} = 102.7 \, {\rm MeV}, \quad
 m_{\tau} = 1746 \, {\rm MeV}
\end{equation}
can be reproduced by adjusting the parameters as follows:
\begin{equation}
 |a_e| = 8.537, \quad
 |C_e| = 1197, \quad
 \beta_e = 11.06
\end{equation}
from which the Yukawa coupling constants for the charged lepton sector
can be fixed provided that $|c_e|\approx|C_e|$. The results are included
in Table \ref{YCCvalues}. Fig.\ref{figureyukawa} shows behavior of
the Yukawa coupling constants of charged fermion sectors.

\begin{figure}[htbp]
\begin{center}
\includegraphics[width=0.6\linewidth]{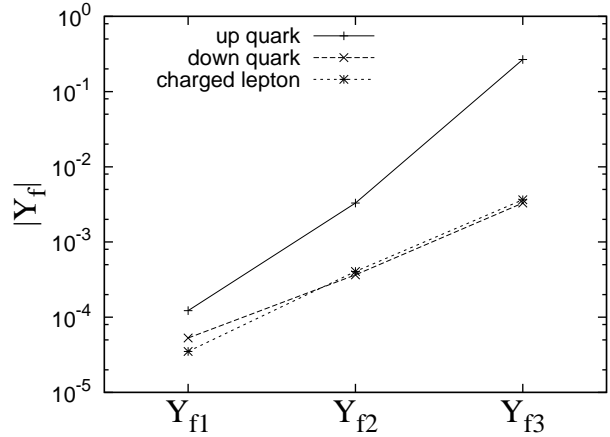}
\end{center}
 \label{figureyukawa}
 \caption{The Yukawa coupling constants $|Y_{f1}|,\,|Y_{f2}|$ and
 $|Y_{f3}|$.}
 \label{figureyukawa}
\end{figure}

\section{Discussion}
We have investigated the non-Hermitian mass matrices $\cM_f$ in
(\ref{EWUpDirac}) and (\ref{EWDownDirac}) deduced in the gauge field theory
of the horizontal symmetry generated by the central extension of the Pauli
algebra. While the matrix $\cM_f$ possesses four complex unknown parameters,
the Hermitian combination $\cM_f\cM_f^\dag$ depends on four real numbers and
two phases. Owing to this fact, we are able to make numerical analyses on
the problem of the six quark masses and the quark FMM with four parameters.

To solve the problem effectively, six real parameters are interpreted as
functions of quark masses and remaining two parameters. Using the observed
masses as inputs and adjusting values of two independent parameters and two
phases, we have reproduced the observed FMM and succeeded to calculate the
Yukawa coupling constants. By careful examination of the numerical results,
empirical relations in (\ref{PhenoRelationYukawa}) are found among the coupling
constants for the quark sector. It is beyond the scope of the present theory
to elucidate physical implications of those relations.

The estimated values of the Yukawa coupling constants in Table
\ref{YCCvalues} show novel orderings with sizable amount of variation.
It should be noticed that the largest ratio between the coupling constants,
${|Y_{u3}|}/{|Y_{u1}|}\sim{|Y_{u4}|}/{|Y_{u1}|}\sim10^3$, is considerably
smaller than that between the observed quark masses ${m_{t}}/{m_{u}}\sim10^5$.
This means that varieties observed directly in the low energy flavor physics
are much reduced at the level of the coupling constants of the Yukawa
interactions.

All results of numerical analyses so far were obtained in the hierarchical
approximation ($m_3^{(f)2} \gg m_1^{(f)2}, m_2^{(f)2}$) by using the
formulae in the Appendix which express $|b_{f1}|, |b_{f2}|$ and $|C_f|$
as functions of $|a_f|^2$. It should be noticed, however, that we are able
to find almost the same results by adjusting the values of real quantities
$|b_{f1}|$, $|b_{f2}|$, $|C_f|$ and $|a_f|^2$ and two phases $\mu$ and $\nu$.

The squared masses of the charged fermions are derived in the exact formula
in (\ref{exactmassfomulae}). Depending on the sign of the angle $\theta$,
the formula has the solutions with normal and partly-reversed orderings
in (\ref{normal}) and (\ref{reversed}). As confirmed in \S \ref{Massordering},
we have to describe the down and up quark states by the normal and
partly-reversed solutions, respectively, in order to reproduce
the observed FMM. This is one of unique features of the present formalism
for the horizontal gauge symmetry. 

We made a speculative analysis of the charged lepton masses by postulating
hypothetical relations (\ref{approximateassmplepton}) as analogues of the
equations for the down quark sector in (\ref{approximateassmpquark}).
To proceed a full investigation on the charged and neutral lepton
sectors, however, it is required to solve the combined eigenvalue problem
which involves not only the Dirac mass matrices but also the Majorana mass
matrix deduced for the neutrino sector.\cite{Sogami.2009}
We will study this problem in near future.

\appendix

\section{Dependence of $|b_1|$, $|b_2|$ and $|C|$ on $|a|$}
In this scheme, we interpret the real quantities $|b_1|$, $|b_2|$ and $|C|$ 
as functions of the quark masses and the parameter $|a|$. For its purpose,
it is appropriate to define a mass $M$ for reference by the relation
\begin{equation}
  m_3^2 + m_2^2 + m_1^2 = |C|^2 + 2|b|^2 + 2|a|^2 = 3M^2.
\label{M.def}
\end{equation}
Evidently observed data in (\ref{quarkmassdata}) show hierarchical orderings
of squared masses, $M^2,\,m_3^2 \gg m_2^2,\,m_1^2$, for each sector.
To deduce convenient relations for numerical analyses, we express all
quantities as power series of $M^2$.

In the hierarchical limit, $P$ and $|C|^2$ take very large values.
Equation (\ref{rhotan3theta}) shows that $(\tan3|\th|)^2$ can be decomposed
in the power series of $M^{-2}$, leading to the approximate relation
\begin{equation}
  (3\th)^2
   \simeq 4 \frac{|a b_1 b_2|^2}{M^6}
   + \frac{1}{3}\frac{\left(|a|^2|b|^2 - 2|b_1 b_2|^2\right)^2}{M^8} .
\label{theta.relation}
\end{equation}
Similarly, the quantity $\delta$ defined by (\ref{delta.def}) is approximated by
\begin{equation}
  \delta \simeq \frac{2}{3}\frac{|a|^2|b|^2 - 2|b_1 b_2|^2}{M^4} .
\label{delta.relation}
\end{equation}
Decomposition of the eigenvalues $m_1^2$ and $m_2^2$ in (\ref{exactmassfomulae})
with respect to $M^{-2}$ results in the following expressions 
\begin{equation}
  m_2^2 + m_1^2 \simeq 2|a|^2 -M^2\delta, \quad
  m_2^2 - m_1^2 \simeq 2\sqrt{3}M^2\th .
\end{equation}
Eliminating $\theta$ and $\delta$ from these equations, we obtain
the relations
\begin{equation}
 |b_1 b_2|^2 = \frac{3}{4|a|^2}(m_2^2 - |a|^2)(|a|^2 -m_1^2)M^2,\quad
 |b|^2=\frac{3}{2}\frac{|a|^4-m_1^2m_2^2}{|a|^4}M^2.
\label{b1b2.b2.relation}
\end{equation}

Then, substitution of the last expression for $|b|^2$ into the defining equation
(\ref{M.def}) allows to express $|C|^2$ in the form
\begin{equation}
 |C|^2=3\frac{m_1^2m_2^2}{|a|^4}M^2-2|a|^2 .
\end{equation}
Finally, to distinguish between $|b_1|$ and $|b_2|$, we introduce
an extra quantity $\kappa$ by
\begin{equation}
 |b_1|^2 = \frac{1}{2}|b|^2-M\kappa, \quad
 |b_2|^2 = \frac{1}{2}|b|^2+M\kappa .
\end{equation}
From (\ref{b1b2.b2.relation}), $\kappa$ is calculated
to be
\begin{equation}
 \kappa = \sqrt{\frac{3}{4|a|^2}
          \left(\frac{3}{4}\frac{(|a|^4-m_2^2m_1^2)^2}{|a|^6}M^2
          - (m_2^2-|a|^2)(|a|^2 - m_1^2)\right)} .
\end{equation}

Consequently, the quantities $|C|$, $|b_1|$ and $|b_2|$ are determined
as the functions of quark masses and the parameter $|a|$ for each sector.
The quark FMM is now expressible in terms of the observed mass values and
the four adjustable parameters $|a_u|$, $|a_d|$, $\mu$ and $\nu$.


\begin{thebibliography}{99}
%%%%%%%%%%%%%%%%%%%%%%%%%%%%%%%%%%%%%%%%%%%%%%%%%%%%%%%%%%%%%
% Some macros are available for the bibliography:
%   o for general use
%      \JL : general journals          \andvol : Vol (Year) Page
%   o for individual journal 
%      \PR  : Phys. Rev.               \PRL : Phys. Rev. Lett.
%      \NP  : Nucl. Phys.              \PL  : Phys. Lett.
%      \JMP : J. Math. Phys.           \CMP : Commun. Math. Phys.
%      \PTP : Prog. Theor. Phys.       \JPSJ: J. Phys. Soc. Jpn.
%      \JP  : J. of Phys.              \NC  : Nouvo Cim.
%      \IJMP: Int. J. Mod. Phys.       \ANN : Ann. of Phys.
% Usage:
%   \PR{D45,1990,345}            ==> Phys.~Rev.\ {\bf D45} (1990), 345
%   \JL{Phys.~Lett.,A30,1981,56} ==> Phys.~Lett.\ {\bf A30} (1981), 56
%   \andvol{B123,1995,1020}      ==> {\bf B123} (1995), 1020
%%%%%%%%%%%%%%%%%%%%%%%%%%%%%%%%%%%%%%%%%%%%%%%%%%%%%%%%%%%%%
%
%Horizontal gauge symmetry
\bibitem{Wilczek.1979} F.~Wilczek and A.~Zee, Phys. Rev. Lett. {\bf 42}
(1979), 421.
\bibitem{Shaw.1993} D.~S.~Shaw and R.~R.~Volkas, Phys. Rev. D {\bf 47}
(1993), 241.
\bibitem{Sogami.2009} I.~S.~Sogami, Prog, Theor. Phys. {\bf 122}, No. 4 (2009)
(in press).
%\bibitem{KobayashiMaskawa} M.~Kobayashi and T.~Maskawa, Prog. Theor. Phys.
%                      {\bf 49} (1973), 652.
%\bibitem{Cabibbo} N.~Cabbibo, Phys. Rev. Lett. {\bf 10} (1963), 531.
\bibitem{Sogami} I.~S.~Sogami, Prog. Theor. Phys. {\bf 114} (2005), 873;
{\bf 115} (2006), 461.
\bibitem{SogamiKon} I.~S.~Sogami and Y.~Konishi, Prog. Theor. Phys.
{\bf 119} (2008), 339.
%Democratic mass matrices
\bibitem{DemocraticMM} H.~Harari, H.~Haut and J.~Weyers, Phys. Lett. B {\bf 78}
                       (1978), 459;\\
                       Y.~Koide, Phys. Rev. D {\bf 28} (1983), 252; 
                       D {\bf 57} (1998), 4429.
\bibitem{Amsler.2008}
C. Amsler {\it et al}. (Particle Data Group), Phys. Lett. B {\bf 667} (2008), 1.
\bibitem{Djouadi.2008} A.~Djouadi, Phys. Rept. {\bf 457} (2008), 1.
\bibitem{ChengLi} T-P.~Cheng and L-F.~Li, {\it Gauge theory of elementary
particle physics} \\ (Clarendon Press, Oxford, 1986), 358.
\bibitem{Fusaoka.1998}
H. Fusaoka and Y. koide, Phys. Rev. D {\bf 57} (1998), 3986.
\bibitem{Xing.2008}
Z. Z. Xing, H. Zhang and S. Zhou, Phys. Rev. D {\bf 77} (2008), 113016.
\bibitem{Jarlskog.1985}
C. Jarlskog, Phys. Rev. Lett. {\bf 55} (1985), 1039.
\end{thebibliography}
\end{document}